\documentclass[a4paper]{jpconf}
\usepackage{graphicx}
\begin{document}
\title{Modeling the jet quenching, thermal resonance production and hydrodynamical flow in relativistic heavy ion collisions}

\author{I P Lokhtin$^{1,*}$, A V Belyaev$^1$, L V Malinina$^1$, S V Petrushanko$^1$, A~M~Snigirev$^1$, I Arsene$^2$ and E E Zabrodin$^{3,1}$}

\address{$^1$ D.V. Skobeltsyn Institute of Nuclear Physics, M.V. Lomonosov Moscow
	State University, Moscow, Russia}

\address{$^2$ Extreme Matter Institute EMMI, GSI Helmholtzzentrum fur Schwerionenforchung GmbH, Darmstadt, Germany}

\address{$^3$ The Department of Physics, University of Oslo, Norway}

\ead{$^*$ Igor.Lokhtin@cern.ch}

\begin{abstract}

The event topology in relativistic heavy ion collisions is determined by various multi-particle production mechanisms. The simultaneous model treatment of different collective nuclear effects at high energies (such as a hard multi-parton fragmentation in hot QCD-matter, thermal resonance production, hydrodynamical flows, etc.) is actual but rather complicated task. We discuss the simulation of the above effects by means of Monte-Carlo model HYDJET++. 
\end{abstract}

\section{Introduction} 

Ongoing and future experimental studies of relativistic heavy ion collisions 
in a wide range of beam energies require the development of new Monte-Carlo 
(MC) event generators and improvement of existing ones. A realistic MC event generator should  
include a maximum possible number of observable physical effects which are 
important to determine the event topology: from the bulk properties of 
soft hadroproduction (domain of low transverse momenta $p_T < 1\div 2$ GeV$/c$)  
such as thermal resonance production and collective flows, to hard multi-parton fragmentation 
in hot and dense QCD-matter, which reveals itself in the spectra of high-$p_T$ particles and 
hadronic jets. HYDJET++ event generator~\cite{Lokhtin:2008xi} includes detailed 
treatment of soft hadroproduction as well as hard multi-parton production, and 
takes into account medium-induced parton rescattering and energy loss. The 
heavy ion event in HYDJET++ is the superposition of two independent components: 
the soft, hydro-type state and the hard state resulting from multi-parton 
fragmentation. Note that a conceptually similar approximation has been 
developed in~\cite{Hirano:2004rs,Armesto:2009zi}. HYDJET++ model is the development and 
continuation of HYDJET event generator~\cite{Lokhtin:2005px}, and it  
contains the important additional features for the soft component: resonance decays and 
more detailed treatment of thermal and chemical freeze-out 
hypersurfaces~\cite{Amelin:2006qe,Amelin:2007ic}. The details on physics model and simulation procedure can be found in HYDJET++ manual~\cite{Lokhtin:2008xi}, the main features of the model being listed only very briefly below.

\section{HYDJET++ model}

The model for the hard multi-parton part of HYDJET++ event is the same as that 
for HYDJET event generator, and it is based on PYQUEN partonic energy loss 
model~\cite{Lokhtin:2005px}. The approach to the 
description of multiple scattering 
of hard partons in the dense QCD-matter (such as quark-gluon plasma) is based on the 
accumulative energy loss via  the gluon radiation being associated with each parton 
scattering in the expanding quark-gluon fluid and includes the interference effect 
(for the emission of gluons with a finite formation time) using the modified radiation 
spectrum $dE/dl$ as a function of decreasing temperature $T$. The model takes into
account radiative and collisional energy loss of hard partons in longitudinally
expanding quark-gluon fluid, as well as realistic nuclear geometry. The event generator for single hard nucleon-nucleon sub-collision PYQUEN was constructed as a modification of the jet event 
obtained with the generator of hadron-hadron interactions PYTHIA$\_$6.4~\cite{pythia}. 
The event-by-event simulation procedure in PYQUEN includes {\it 1)} generation of 
initial parton spectra with PYTHIA and production vertexes at given impact parameter; 
{\it 2)} rescattering-by-rescattering simulation of the parton path in a dense zone 
and its radiative and collisional energy loss; {\it 3)} final hadronization according 
to the Lund string model for hard partons and in-medium emitted gluons. Then the 
PYQUEN multi-jets generated according to the binomial distribution are included in the 
hard part of the event. The mean number of jets produced in an AA event is the 
product of the number of binary NN sub-collisions at a given impact parameter and the 
integral cross section of the hard process in $NN$ collisions with the minimum  
transverse momentum transfer $p_T^{\rm min}$. In order to take into account the 
effect of nuclear shadowing on parton distribution functions, the impact parameter 
dependent parameterization obtained in the framework of Glauber-Gribov 
theory~\cite{Tywoniuk:2007xy} is used. 

The soft part of HYDJET++ event is the ``thermal'' hadronic state generated on the 
chemical and thermal freeze-out hypersurfaces obtained from the parametrization 
of relativistic hydrodynamics with preset freeze-out conditions (the adapted C++ code 
FAST MC~\cite{Amelin:2006qe,Amelin:2007ic}). Hadron multiplicities are calculated 
using the effective thermal volume approximation and Poisson multiplicity distribution 
around its mean value, which is supposed to be proportional to the number of 
participating nucleons at a given impact parameter of AA collision. The fast soft 
hadron simulation procedure includes {\it 1)} generation of the 4-momentum of a hadron 
in the rest frame of a liquid element in accordance with the equilibrium distribution 
function; {\it 2)} generation of the spatial position of a liquid element and its 
local 4-velocity in accordance with phase space and the character of motion of the 
fluid; {\it 3)} the standard von Neumann rejection/acceptance procedure to account 
for the difference between the true and generated probabilities; {\it 4)} boost of 
the hadron 4-momentum in the center of mass frame of the event; {\it 5)} the two- 
and three-body decays of resonances with branching ratios taken from the SHARE 
particle decay table~\cite{share}. The high generation speed in HYDJET++ is achieved 
due to almost 100\% generation efficiency of the ``soft'' part because of the  
nearly uniform residual invariant weights which appear in the freeze-out 
momentum and coordinate simulation. 

Note that although HYDJET++ is optimized for very high energies of RHIC and LHC colliders (c.m.s. energies of heavy ion beams $\sqrt{s}=200$ and $2760\div 5500$ GeV per nucleon pair 
respectively), it can also be used for studying the  
particle production in a wider energy range down to $\sqrt{s} \sim 10$ GeV per nucleon pair at future facilities FAIR and NICA. As one moves from 
very high to moderately high energies, the contribution of the hard part of the event 
becomes smaller, while the soft part turns into just a multi-parameter fit to the data.

\section{Some applications of HYDJET++ at RHIC and LHC}

It was demonstrated in~\cite{Lokhtin:2008xi} that HYDJET++ model can 
describe the bulk properties of hadronic state created in Au+Au collisions 
at RHIC at $\sqrt{s}=200 A$ GeV (hadron spectra and ratios, radial 
and elliptic flow, femtoscopic momentum correlations), as well as the high-$p_T$ hadron 
spectra. A number of input parameters of the model can be fixed from fitting 
the RHIC data to various physical observables. For example, the thermodynamical
potentials and the chemical freeze-out temperature  $T^{\rm ch}=0.165$ GeV have been fixed 
in HYDJET++ from fitting the RHIC data to hadron ratios near mid-rapidity in central 
Au+Au collisions. The slopes of transverse mass $m_T$ hadron spectra ($\pi^+$, $K^+$ and $p$ with $m_T<0.7$ GeV/$c^2$) near mid-rapidity at different centralities 
spectra allow the thermal freeze-out temperature  $T^{\rm th}=0.1$ GeV and 
the maximal radial flow rapidity in central collisions 
$\rho_{\rm max}(b=0)=1.1$ to be fixed. The space-time parameters of thermal 
freeze-out region can be fixed by means of fitting the three-dimensional correlation 
functions measured for $\pi^+\pi^+$ pairs and extracting the correlation radii 
$R_{\rm side}$, $R_{\rm out}$ and $R_{\rm long}$. Rapidity spectra of 
charged hadrons at different centralities allow us to fix the particle densities in the 
mid-rapidity region and the maximum longitudinal flow rapidity 
$\eta_{\rm max}=3.3$. Since mean ``soft'' and ``hard''
hadron multiplicities depend on the centrality in different ways, the relative contribution of 
soft and hard parts to the total event multiplicity can be fixed through the 
centrality dependence of $dN/d\eta$. High transverse momentum hadron 
spectra ($p_T > 2\div 4$ GeV/$c$) are sensitive to parton production and 
jet quenching effect, therefore fitting the measured high-$p_T$ tail allows 
the extraction of PYQUEN energy loss model parameters. The momentum and azimuthal 
anisotropy parameters are estimated for different centrality sets by 
fitting the measured transverse momentum dependence of the elliptic flow coefficient, $v_2(p_T)$. 

\begin{figure}[h]
\begin{minipage}{38pc}
\includegraphics[width=18pc]{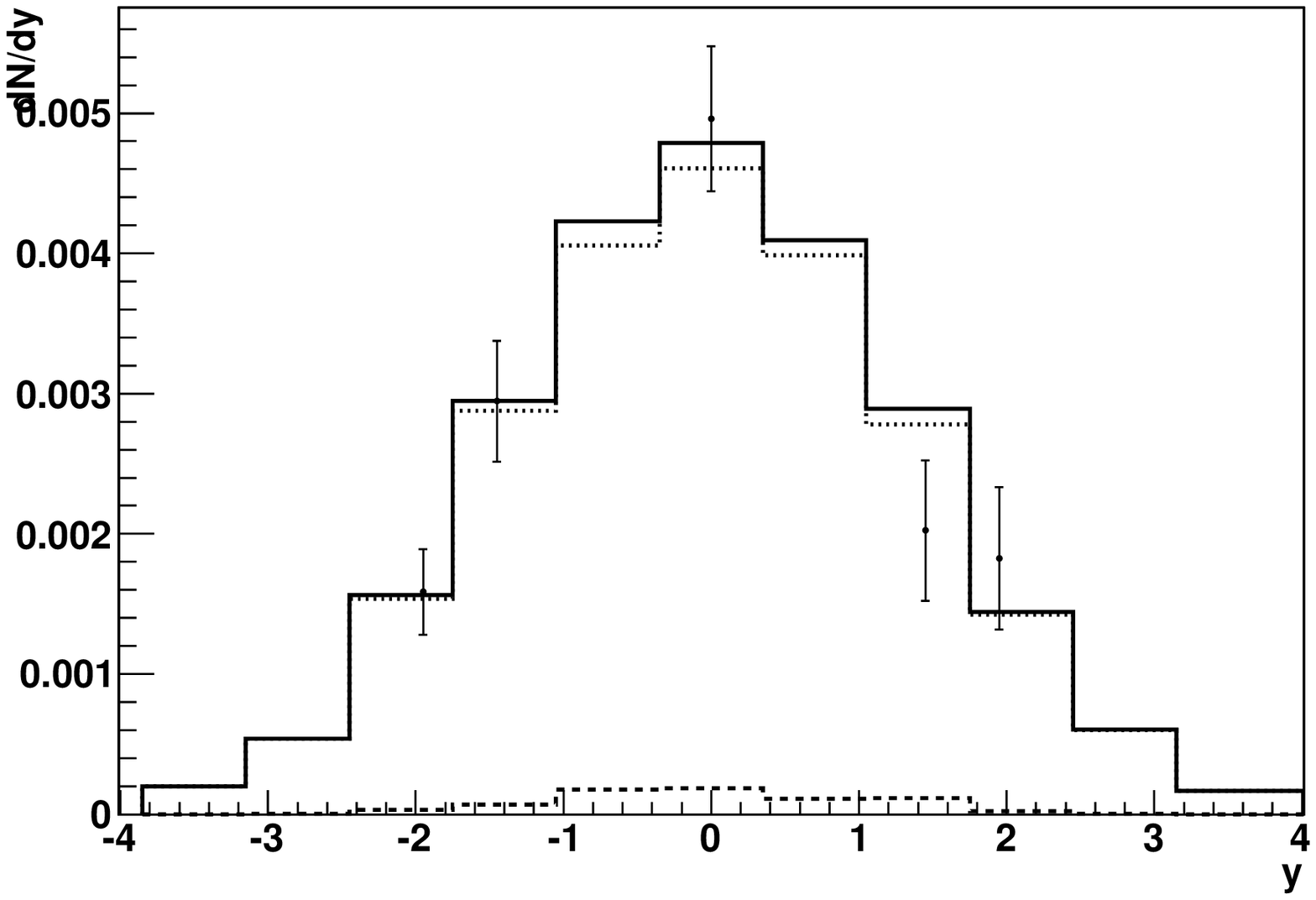}
\hspace{2pc}
\includegraphics[width=18pc]{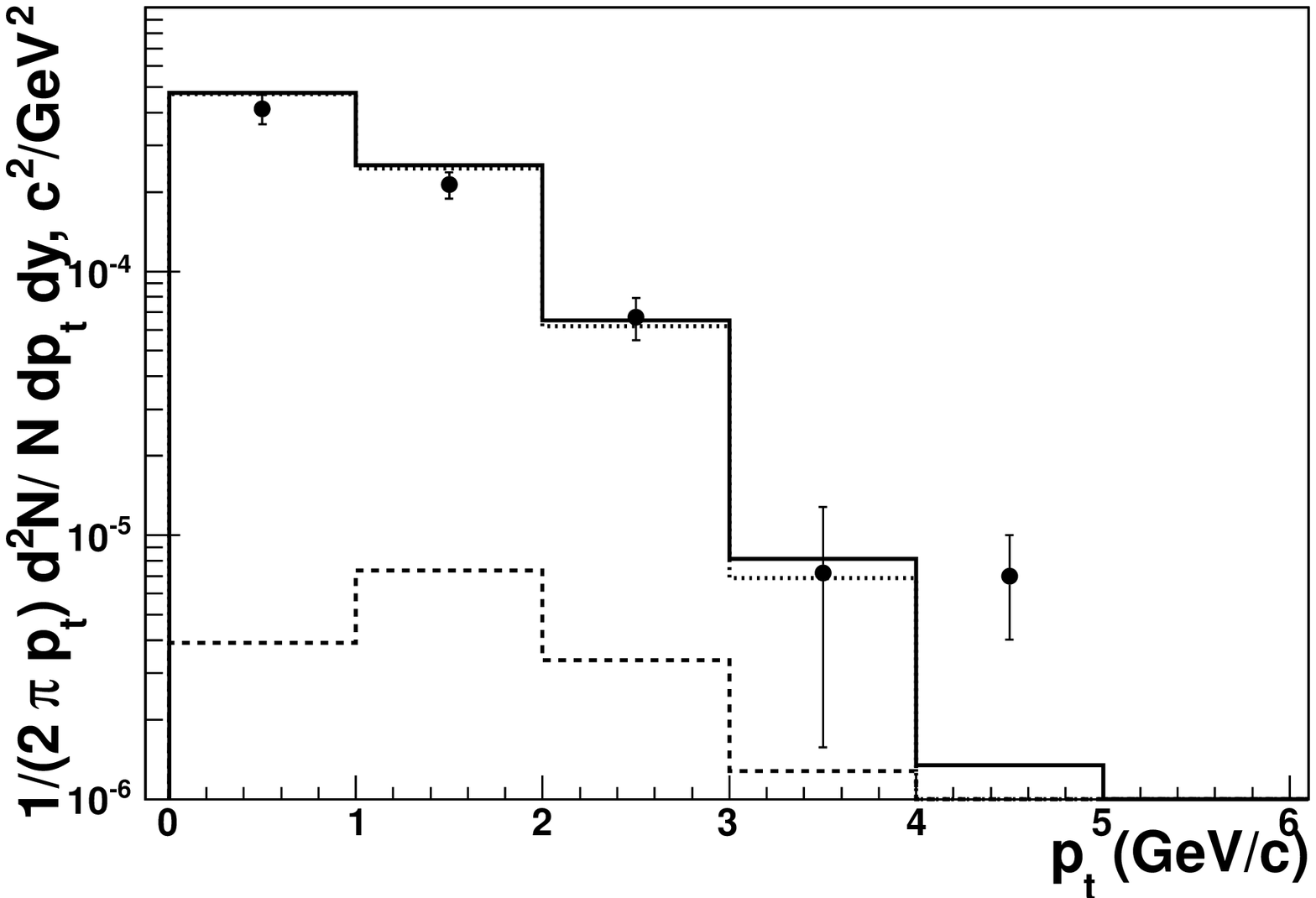}
\caption{\label{fig} The pseudorapidity (left) and the transverse 
momentum (right) spectra of $J/\psi-$mesons in Au+Au collisions at $\sqrt{s}=200 A$ GeV for 
$0\div 20$\% centrality. The points are PHENIX data~\cite{Adare:2006ns}, histograms are the HYDJET++ calculations for  
$\gamma_c=7$ and $T^{\rm th}(J/\psi)=T^{\rm ch}=0.165$ GeV (solid -- total, dotted -- soft component, dashed -- hard component).}
\end{minipage} 
\end{figure}

Thermal charm hadron production was implemented in HYDJET++ recently. $D$, $J/\psi$ and $\Lambda_c$ hadrons are generated within the statistical hadronization model~\cite{Andronic:2003zv,Andronic:2006ky}. Momentum spectra of charm hadrons are computed according to the thermal distribution, and the multiplicities $N_c$ ($C = D, J/\psi, \Lambda_c$) are calculated through the corresponding thermal numbers $N_c^{\rm th}$ as 
\begin{equation}
N_c=\gamma_c^{\rm nc}  N_c^{\rm th}~,
\end{equation}
where $\gamma_c$ is the charm enhancement factor (or charm fugacity), and $N_c$ is the number of charm quarks in a hadron $C$. The fugacity $\gamma_c$ can be treated as a free parameter of the model or  calculated through the number of charm quark pairs obtained from perturbative QCD. HYDJET++ fits measured by PHENIX $J/\psi$ yields in central Au+Au collisions~\cite{Adare:2006ns} with $\gamma_c=7$. If we assume that thermal freeze-out for $J/\psi$-mesons happens at the same temperature as for light hadrons, $T^{\rm th}=0.1$ GeV ($\eta_{\rm max}=3.3$, $\rho_{\rm max}=1.1$), then simulated $y-$ and $p_T-$spectra are much wider than the data. However if we assume that thermal freeze-out for $J/\psi$-mesons occurs at the same temperature as chemical freeze-out, $T^{\rm th}(J/\psi)=T^{\rm ch}=0.165$ GeV, then simulated spectra are capable of fitting the PHENIX data for the values of maximal longitudinal and radial flow rapidities 
$\eta_{\rm max}^{\rm ch}=1.1$ and $\rho_{\rm max}^{\rm ch}=0.5$ respectively (see Fig.~\ref{fig}). 

The heavy ion collision energy at LHC will be a factor of $\sim 15\div 30$ larger then that in RHIC, thereby allows one to probe new frontiers of super-high temperature and (almost) net-baryon free QCD. 
It is expected that at such ultra-high energies the role of hard and semi-hard particle production 
may be significant even for the bulk properties of created matter. In particular, the spectacular 
predictions of HYDJET++ are possible reducing the femtoscopic correlation radii~\cite{Lokhtin:2009be} and elliptic flow~\cite{Eyyubova:2009hh} in heavy ion collisions as one moves from RHIC to LHC energies due to the significant contribution of (semi)hard component to the space-time structure of the hadron emission source. 

\section{Summary} 

Among other heavy ion event generators, Monte-Carlo model HYDJET++ focuses on the detailed 
simulation of jet quenching effect basing on the partonic energy loss model 
PYQUEN, and also reproducing the main features of nuclear collective dynamics 
by the parametrization of relativistic hydrodynamics with preset freeze-out conditions (including 
resonance decays, and separate treatment of thermal and chemical freeze-out hypersurfaces). 
Thus the final hadronic state in HYDJET++ represents the superposition of two independent components: medium-modified hard multi-parton fragmentation and soft hydro-type part. HYDJET++ is capable of reproducing the bulk properties of heavy ion collisions at RHIC (hadron spectra and ratios, radial 
and elliptic flow, femtoscopic momentum correlations), as well as hard probes (high-p$_T$ 
hadron and $J/\psi$ spectra). The simulations for LHC (and also at lower energies of future facilities 
FAIR and NICA) are in progress. 

\section*{Acknowledgments}  

We would like to thank L.V.~Bravina, D.~d'Enterria, G.Kh.~Eyyubova, A.M.~Gribushin,  V.L.~Korotkikh,  R.~Lednicky, C.~Loizides, L.I.~Sarycheva,  Yu.M.~Sinyukov and K.~Tywoniuk for joint work and fruitfull discussions. I.L. and S.P. wish to express the gratitude to the organizers of the Workshop ``Hot Quarks 2010'' for the warm welcome and the hospitality. This work was supported by Russian Foundation for Basic Research (grants Nos 08-02-91001, 08-02-92496 and 10-02-93118), Russian Ministry for Education and Science (contract 02.740.11.0244) and Dynasty Foundation.

\end{document}